\documentclass[sigconf]{acmart}

\usepackage{subcaption}
\usepackage{balance}

\newcommand{\IT}[0]{\textit{{TeachNow}}}

\def\colorful{1}

\ifnum\colorful=1

\newcommand{\red}[1]{{\color{red} {#1}}}

\fi
\ifnum\colorful=0

\newcommand{\red}[1]{{{#1}}}

\fi

\AtBeginDocument{%
  \providecommand\BibTeX{{%
    \normalfont B\kern-0.5em{\scshape i\kern-0.25em b}\kern-0.8em\TeX}}}

\copyrightyear{2024}
\acmYear{2024}
\setcopyright{acmlicensed}\acmConference[ITiCSE 2024]{Proceedings of the 2024 Innovation and Technology in Computer Science Education V. 1}{July 8--10, 2024}{Milan, Italy}
\acmBooktitle{Proceedings of the 2024 Innovation and Technology in Computer Science Education V. 1 (ITiCSE 2024), July 8--10, 2024, Milan, Italy}
\acmDOI{10.1145/3649217.3653629}
\acmISBN{979-8-4007-0600-4/24/07}

\begin{document}



\title{\IT{}: Enabling Teachers to Provide Spontaneous, Realtime 1:1 Help in Massive Online Courses}







\author{Ali Malik}
\authornote{Both authors contributed equally to this research.}
\email{malikali@cs.stanford.edu}
\orcid{0009-0007-1201-5014}
\affiliation{%
  \institution{Stanford University}
  \state{California}
  \country{USA}
}

\author{Juliette Woodrow}
\authornotemark[1]
\email{jwoodrow@stanford.edu}
\orcid{0009-0006-8097-093X}
\affiliation{%
  \institution{Stanford University}
  \state{California}
  \country{USA}
}

\author{Chao Wang}
\email{20cw50@queensu.ca}
\affiliation{%
  \institution{Queen's University}
  \city{Ontario}
  \country{Canada}
}
\author{Chris Piech}
\email{piech@cs.stanford.edu}
\orcid{0000-0001-5140-0467}
\affiliation{%
  \institution{Stanford University}
  \state{California}
  \country{USA}
}





\renewcommand{\shortauthors}{Malik and Woodrow et al.}

\begin{abstract}
    One-on-one help from a teacher is highly impactful for students, yet extremely challenging to support in massive online courses (MOOCs). 
    In this work, we present \IT{}: a novel system that lets volunteer teachers from anywhere in the world instantly provide 1:1 help sessions to students in MOOCs, without any scheduling or coordination overhead. \IT{} works by quickly finding an online student to help and putting them in a collaborative working session with the teacher. 
    The spontaneous, on-demand nature of \IT{} gives teachers the flexibility to help whenever their schedule allows.
    
    We share our experiences deploying \IT{} as an experimental feature in a six week online CS1 course with 9,000 students and 600 volunteer teachers. Even as an optional activity, \IT{} was used by teachers to provide over 12,300 minutes of 1:1 help to 375 unique students. 
    Through a carefully designed randomised control trial, we show that \IT{} sessions increased student course retention rate by almost 15\%. Moreover, the flexibility of our system captured valuable volunteer time that would otherwise go to waste.
    Lastly, \IT{} was rated by teachers as one of the most enjoyable and impactful aspects of their involvement in the course. We believe \IT{} is an important step towards providing more human-centered support in massive online courses.
    

    

\end{abstract}





\maketitle

\begin{figure*}
    \centering
    \includegraphics[width=0.95\linewidth]{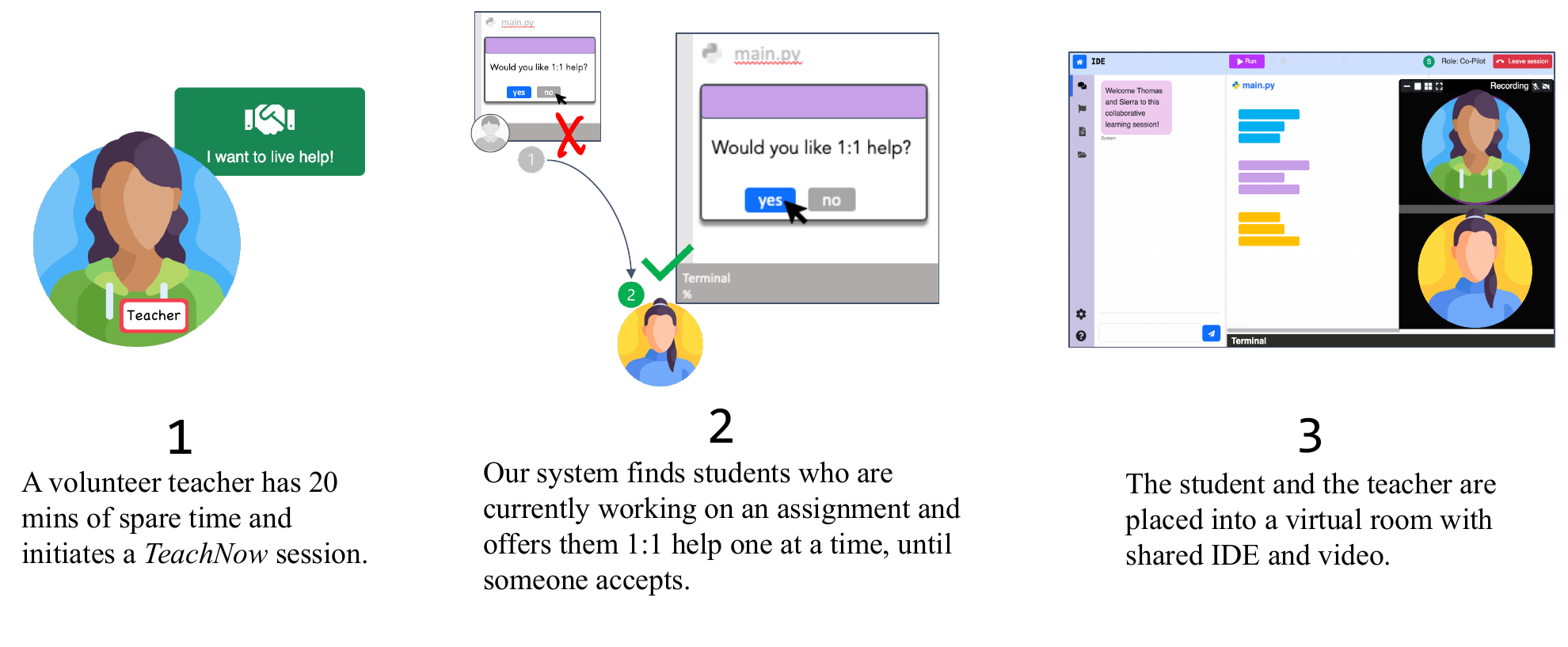}
    \caption{A volunteer teacher initiates the \IT{} process whenever they have spare time. She presses a button and the system finds available students working on assignments at that moment. Students are nudged one by one with an offer for a 1:1 session with the volunteer until a student accepts. It takes less than 5 minutes for a match to be found. Once the student accepts, the student and volunteer teacher are put into a virtual 1:1 session.}
    \label{fig:overview}
\end{figure*}

\section{Introduction}

One-on-one live help from a teacher is one of the most powerful and authentic learning experiences. Research has shown that this type of interaction has profound impacts on student understanding \cite{Bloom82TwoSigma, carter2009visible, ericsson1993role}, increased identity formation with the subject, and a deeper sense of belonging \cite{piech2021code, dweck2014academic}. 

However, scaling this kind of support to tens-of-thousands of students in massive online courses (MOOCs) has always seemed daunting. The obvious problem is finding enough teachers. Fortunately, notable progress has been made on this front by leveraging online volunteer teachers \cite{PythonTutor, piech2021code}. 
For example, MOOCs like Code in Place consistently recruit thousands of volunteer teachers each year to take on supportive roles such as teaching group section or answering discussion forum questions \cite{piech2021code}.
Could these volunteers also be used to provide 1:1 help to students of the course?

\red{
}

\subsection{The challenges of 1:1 help at scale}
While volunteers solve the scale problem, there remain several key challenges in leveraging them for 1:1 help.

\paragraph{\textbf{Complex coordination.}}
Volunteer teachers are a decentralised workforce, 
potentially distributed across 100+ countries and 24 timezones.
The students in massive online courses are similarly dispersed. This makes even the simple task of coordinating a time for 1:1 help difficult. Any time wasted in coordinating a meetup foregoes valuable minutes that could be used helping a student.

\paragraph{\textbf{Teacher time is sporadic and unpredictable.}}
Much of the free time volunteers have to help students is scattered throughout the week, sometimes in fragmented intervals of 20 minutes. Approaches for 1:1 help that require pre-scheduling sessions often fail to capture these valuable chunks of time.


\paragraph{\textbf{Pitfalls of students seeking help}} In existing approaches to 1:1 help, such as office hours, students are responsible for seeking out help. This requires the motivation to write down their confusion, find a scheduled time, and wait in a queue for help. The moment they receive help could be days after losing the mental context of their original problem \cite{hines2004increasing}. 
Moreover, long wait times and overcrowded office hours can make students more frustrated.

Compounding this issue, constant, readily available help can cultivate a dependency known as ``learned helplessness'' \cite{Dweck1973LearnedHA}  – a habitual reliance on seeking help that can impede the learning process. Learned helplessness may condition students to avoid attempting tasks independently, leading them to depend heavily on these office hours for problem-solving assistance instead of tackling the work by themselves.



\subsection{Our key insight}

How can we design a system that enables volunteer teachers to effectively provide 1:1 help in MOOCs? Such a system should makes optimal use of teacher availability, avoid overpromising and disappointing students, and provide  in-context help to students when they are actively doing work.

Our key insight is to move away from a model where students seek help during allotted times (demand-side), to instead \textbf{giving teachers the ability to initiate 1:1 help}
(supply-side). 

\begin{quote}
\textbf{What if a teacher could offer 1:1 help at any moment, whenever their schedule allowed for it, and we could quickly find a student for them to help?    }
\end{quote}





In this paper, we present \IT{}: an on-demand system that let's a volunteer teacher instantly provide 1:1 help to a student in a massive online course. When a teacher initiates a \IT{} session, our tool looks for a student who is actively working on an assignment and offers them 1:1 help. If the student declines, our system tries another person. When a student accepts, the teacher and student are automatically placed into a shared virtual space with live
video/audio/chat and a shared editor to collaborate on (Figure \ref{fig:overview}).

\IT{} gives teachers the flexibility to provide 1:1 help at any time, without any scheduling or coordination overhead. This allows them to channel fragmented moments of their time into helping students.

\subsection{Main contributions}

We highlight the main contributions of our paper.

\noindent \quad \textbf{(1)}  We  implement a  technically challenging, production-level system that enables instantaneous, teacher-initiated 1:1 help session in massive online courses. \IT{} could be integrated as a general feature in most massive online courses (Section \ref{sec:teachnow}).

\noindent \quad \textbf{(2)} We deploy \IT{} in a real MOOC with 9,000 students and 600 volunteer teachers from across the world. Even as an optional feature, our system was used to provide 12,300 minutes of 1:1 help to over 375 unique students (Section \ref{sec:deployment}).

\noindent \quad \textbf{(3)} Through a carefully design randomised control trial (RCT), we show that a single 1:1 help session increases student course retention rates by 15\%. Moreover, we find strong evidence that the teacher-initiated aspect of \IT{} captures valuable volunteer time that would otherwise go to waste (Section \ref{sec:results}).


\noindent \quad \textbf{(4)} Through qualitative surveys and interviews, we show that both students and teachers  find \IT{} to be one of the most gratifying and impactful aspects of the course (Section \ref{sec:results}).


\section{Background and Related Work}\label{sec:related_work}



\paragraph{\textbf{One-on-one help}} 
Receiving personalised, 1:1 help from an expert is known to be one of the most effective educational interventions a student can receive \cite{piech2021code, dweck2014academic,carter2009visible, ericsson1993role}. As early as the 80's, \citet{Bloom82TwoSigma} showed that this kind of support helped students perform almost two standard deviations above regular students. As a result, providing something akin to 1:1 help at the scale of massive online courses has been a holy grail in education research.  
    
\paragraph{\textbf{AI Tutors}}
Several compelling attempts have been made to tackle this problem. One branch of research leverages advances in Artificial Intelligence (AI) to support students as a human would \cite{AutoTutor, DeepTutor, Guru, SmartMOOC, ExampleTracingTutor, ElectronixTuto, SKOPEIT, ASSISTments, ITLitReview}. While some of these results are powerful and helpful to students, they do not provide human connection to learners and are not as effective as humans at providing this 1:1 support. Notably, they lack the essential socio-psychological factors present in interaction with a human expert that are \textit{crucial} to student learning and motivation \cite{SIT2005140, angelino2007strategies, ali2015comparing}

\paragraph{\textbf{Peers and Volunteer Teachers to Support Learning}}
A recent string of work has rallied under the umbrella of ``human-centered learning at scale'': how can one leverage humans to provide support to learners at scale? Famous explorations in this area include the vast literature on peer teaching \cite{SectionLeading2017, SectionLeading1998, whitman1988peer}, feedback \cite{PeerStudio, PeerSelfAssess, PeerGradeOpenResponse, PeerAssesMOOC}, and collaboration \cite{PeerSupport, ProjectBasedCollabLearning, CreatingCollaborativeGroups, DynamicPeerCollab} in MOOCs.

More recently, investigations have explored creative ways to scale student-teacher interactions in MOOCs. One example, Codeopticon \cite{Codeopticon}, is a tool designed that allows the teacher to effortlessly monitor 12 students at a time and initiate a chat session with any of them at any given time. 

Another example is the Code in Place MOOC  \cite{piech2021code} which enlists 1,000 volunteer teachers who have experience with the course material to teach weekly discussion sections to groups of 10-12 students.

A key insight from these approaches is that the scale problem \textit{can} be solved with people: in other words, the number of qualified people who can support learners is roughly proportional to the number of learners.  
\IT{} capitalizes on this key insight, utilizing volunteer teachers to provide teacher initiated one-on-one assistance to students in large-scale online courses.



\section{\IT{}}
\label{sec:teachnow}

We present \IT{}: a first-of-its-kind tool designed to enable teachers to provide one-on-one help sessions to students whenever the teacher has free time, without any scheduling or coordination overhead. 

We walk through the interface and usage of our tool with a concrete example: 
Sierra is a volunteer teacher in the online course. She finds herself with 20 minutes of free time and wants to use it to help students.


\paragraph{\textbf{Step 1: Teacher initiates a \IT{} session}}


Sierra navigates to the course home page and clicks the the button to initiate a \IT{} session. A waiting screen pops up while she waits for a match.


\paragraph{\textbf{Step 2: Finding a student to help}}

Once a session has been initiated, our system
looks for an appropriate student to offer 1:1 help to.  We look for students who are (1) online within the past minute, (2) actively working on a homework problem in the IDE, and (3) have not been offered help recently in the past 24 hours. 
We call the set of students that satisfy the above criteria the \textbf{\textit{nudgable}} students. From this nudgeable set, our algorithm randomly selects one student for a 1:1 help session. 



\paragraph{\textbf{Step 3: Help offered to student}}

Thomas is the student who is randomly selected. He is working on his assignment in the integrated development environment (IDE) \cite{TJIDE} when he receives a popup notification offering help from a volunteer teacher.


Thomas has 30 seconds to accept or reject the offer, otherwise the nudge is automatically rejected. If Thomas does not want to meet, he can click ``Not now.'' In this case, we go back to Step 2 to find another student to nudge.

We nudge students one at a time to ensure a teacher is always available when a student accepts the offer.
If Thomas agrees to the session, both Sierra and Thomas will receive a notification linking to their 1:1 session page. 



\paragraph{\textbf{Step 4: Virtual 1:1 Help}}


Once they click to join, Sierra and Thomas are each given a chance to set their audio and video preferences. They are then directed to a collaborative space with an IDE showing Thomas's code from the assignment he is working on (Figure \ref{fig:overview}). The interface includes a live video stream and a chat option. Both participants can view and run the code, but only Thomas can make edits. 

Sierra's job is to help the student if they have questions, or check in on his general progress and  experience with the course.

\paragraph{\textbf{Step 5: Post-session}}
After the 1:1 session ends, Thomas is prompted with the option to thank his volunteer teacher. If he wishes, he can also leave a personalised message of appreciation, which we use as a signal on how the student's session went. Note that we did not share the student message text with volunteer section leaders right away as we wanted to verify that everything written and sent was appropriate and kind.

At the end of the session, Sierra is prompted to evaluate the session by rating her experience and providing any additional comments. She also gets to see an aggregated count of the gratitude she has received from her \IT{} sessions.

\section{Deploying \IT{} in  a Massive Online Course}\label{sec:deployment}


We deployed \IT{} in a real massive online course on introductory programming, with around 9,000 students enrolled from all over the globe.
The course ran synchronously for 6 weeks and involved watching lecture videos, doing readings, weekly programming assignments, and a final exam. 
The course also enlisted 600 volunteer teachers who were responsible for teaching a weekly section (1 hour/week) and answering questions on a discussion forum (1 hour/week).  

We deployed \IT{} as an experimental, opt-in feature for volunteers to use from the second week of the course. All volunteer usage of the tool was entirely optional.



\subsection{Overview and Usage}
\begin{figure}[t]
    \centering
    \includegraphics[width=0.99\linewidth]{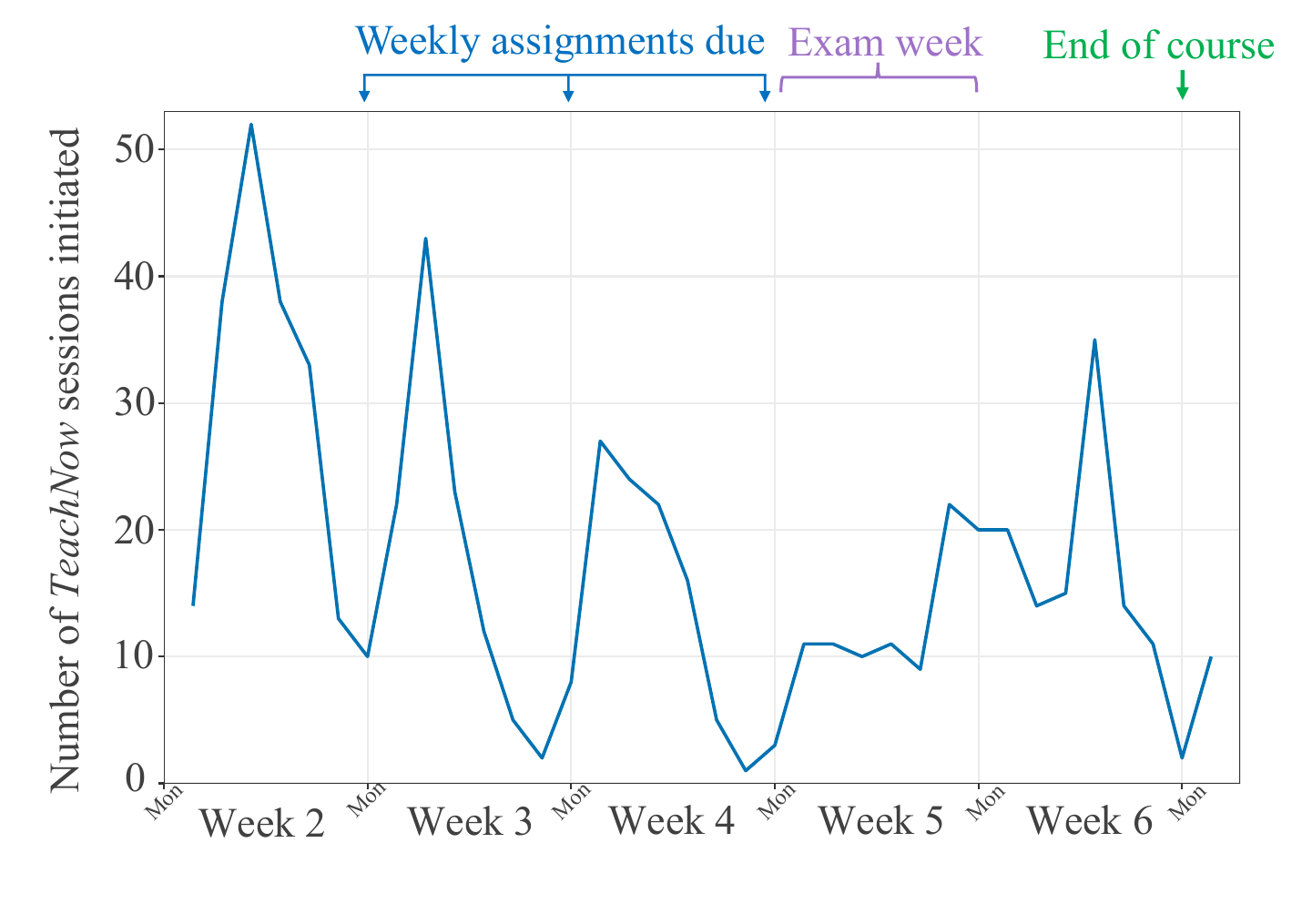}
    \caption{\IT{}  sessions over the weeks in the course, starting from Week 2. }
    \label{fig:tickets_over_time}
\end{figure}

Table \ref{tab:summary} presents a summary of how \IT{} was used by volunteer teachers. Despite being an optional activity, over 15\% of the 600 volunteer teachers used our system to provide a collective 12,300 minutes of 1:1 help to 375 unique students. Notably, 80\% of the volunteer teachers who had a \IT{} session returned to use the tool again. We also found sustained usage of \IT{} throughout the course (Figure \ref{fig:tickets_over_time}), with a clear cyclic usage pattern. 





In a typical \IT{} session, a volunteer teacher would check in with a student to identify any specific learning obstacles, provide personalised assistance with programming assignments, or revisit key concepts from lectures. These sessions also doubled as community forming social interactions, creating opportunities for participants to connect with volunteers from around the globe.



In terms of safety, each \IT{} session was recorded with the explicit consent of participants. 
Additionally, we provided a reporting mechanism for both students and teachers to communicate any incidents. Throughout the course, we did not receive any reports of inappropriate behaviour or negative interactions with \IT{}.

\begin{table}
    \centering
    \begin{tabular}{ll}
        \toprule
        \textbf{Statistic} &  $n$ \\ 
        \midrule 
        \textbf{Volunteer Teachers} &  \\
            Number  who tried \IT{} & $102$ \\
            Number of tickets initiated & $679$ \\
            Number of tickets accepted & $411$ \\
            Median number of tickets per volunteer & $4$ \\
        \\
        \textbf{Students} &  \\ 
           Number of unique students nudged & $1056$ \\
           Number of unique students helped & $375$ \\
        \bottomrule
    \end{tabular}
    \caption{Summary statistics of \IT{} sessions.}
    \label{tab:summary}
\end{table}

\subsection{Example stories of \IT{}}
We share some select examples of \IT{} sessions that occurred during the 6 week course. 

 \textbf{\textit{Thailand. }}
The professor of the course tried \IT{} and matched with a student in Thailand, surprising them with the unique opportunity to directly meet the professor. They discussed the bugs in the student's program and the instructor got to hear first hand from students about how the course was going.

 \textbf{\textit{Brazil. }} One teacher matched with a 32 year old student in Brazil. The student didn't need help but just wanted to talk about the course. The two discovered a shared passion for languages and had a tri-lingual conversation in English, Spanish and Portuguese.

  \textbf{\textit{USA. }} A teacher in California matched with a 45 year old student in New Jersey. Early in the session, the student shared he was inspired to take the course to retrain and pursue a job in CS, but that he was struggling a lot with the course material. In the session, the teacher clarified some fundamental misunderstandings, which helped clear the way for the student to tackle the problem.

\section{How impactful is \IT{}?}\label{sec:results}

Given the large-scale nature of the MOOC setting, we designed several experiments to quantitatively understand the impact of \IT{} on  students and teachers. Each subsection outlines a different finding, along with evidence to support the claim.

\subsection{A single \IT{} session significantly increases student retention.}
In the context of massive online courses, an essential indicator for learning is student progression through the course. MOOCs are notorious for high dropout rates ($\approx 95\%$ by the end of the course) \cite{chuang_ho_2016}, so  strategies to retain students are extremely valuable.


\subsubsection{\textbf{Experimental setup}}

We measured course progress as the percentage of all assignments completed by the learners, where completion was determined using automated unit tests. We also computed ``retention'', which was the fraction of learners who were still active in the course by a certain day.

To measure the causal impact of \IT{} sessions on students, we designed a randomised control trial (RCT). 
From the start of Week 2 to the end of Week 3, only a random subset of $35\%$ of students  were eligible to be nudged for \IT{} help. Everyone else was in a restricted group which would not be offered 1:1 help by our system\footnote{At the end of Week 3, eligibility was extended to the all students.}.
For each \IT{} session that occurred during this time, we carefully constructed a control group via propensity matching. At the exact moment a student accepted help for a \IT{} session, we found a set of ``counterfactual'' students to compare her with. Specifically, we looked at all other users who were nudgable in that instant (i.e online and on the IDE) but happened to be in the random restricted group. From this group, we filtered to students who were working on the same assignment as the helped student and were within $\pm 1\%$ of this student's course progress.
This control strategy ensured we made fair comparisons between each student who was helped and other students in the course.

\subsubsection{\textbf{Results}}

\begin{figure}
    \begin{subfigure}{.99\linewidth}
    \centering
    \includegraphics[width=0.8\linewidth]{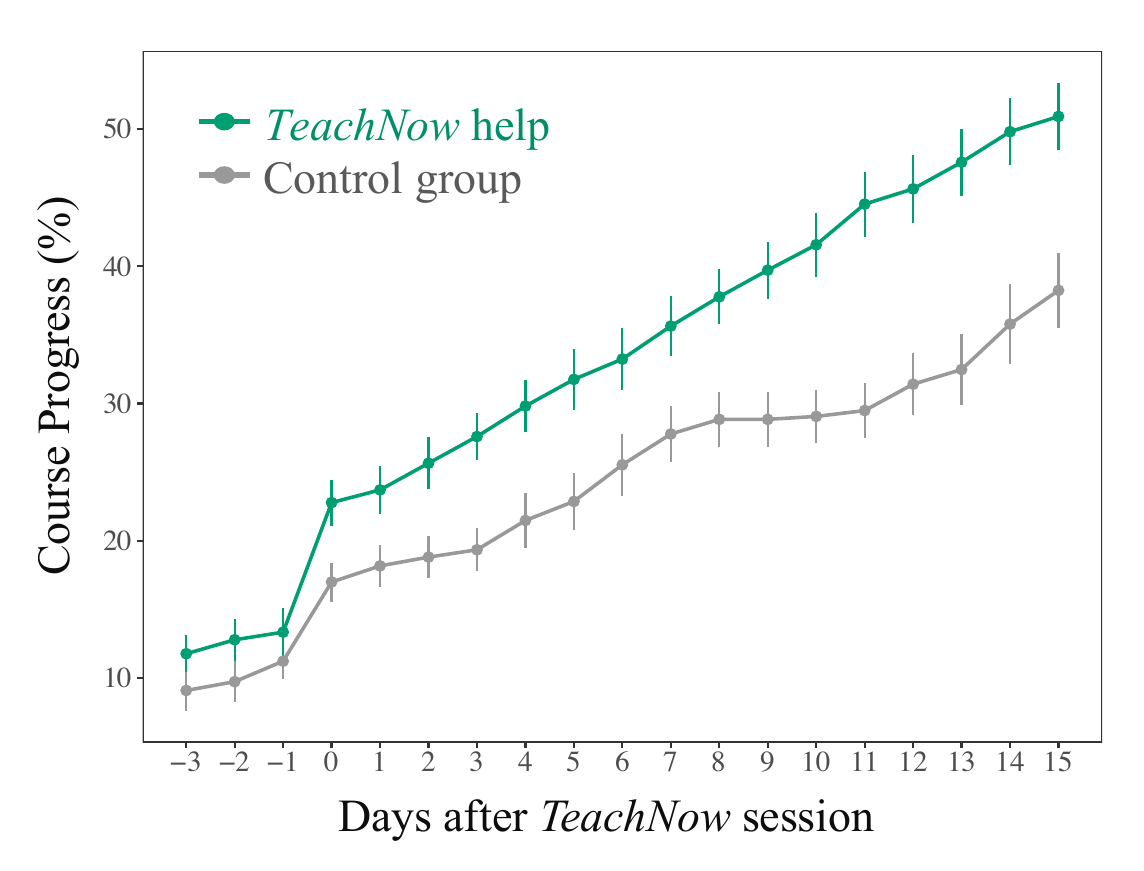}
    \end{subfigure}
    \begin{subfigure}[t]{.99\linewidth}
    \centering
    \includegraphics[width=0.78\linewidth]{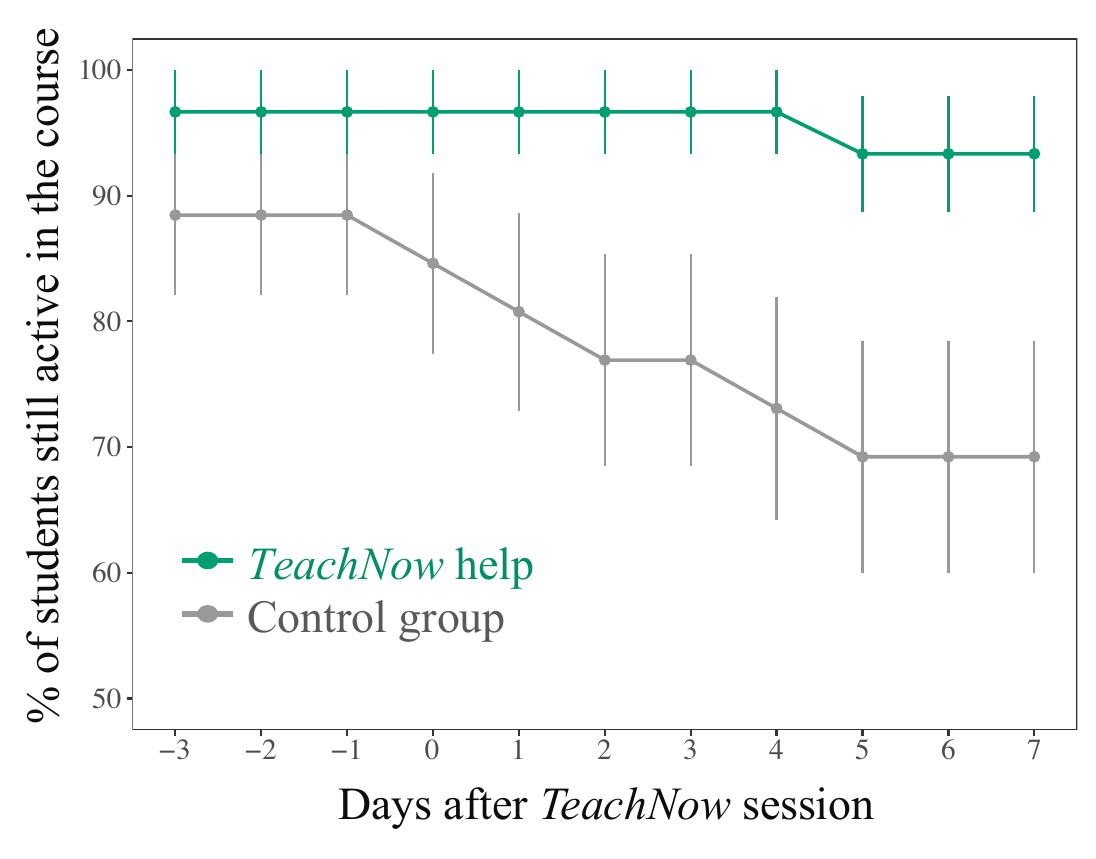}
    \end{subfigure}
    \caption{Student course progress and dropout impacted by a \IT{} session.}
    \label{fig:impact}    
\end{figure}

We compared the average course progress between the students who received \IT{} help and the propensity control group, starting from $3$ days before the session to $15$ days after the session (Figure \ref{fig:impact} (top)).
As expected, on days prior to the session, there was no significant difference. However, within a day of a $\IT{}$ session, the helped group started performing significantly better than the control group in terms of course progress.

These results indicate that students who received a single \IT{} session had a \textbf{\textit{sustained}} 40\% higher course progression than the control group.
Further investigation revealed that the primary cause for this difference was ``dropout'': once a student quit, their stagnant progress would drag down the average course progress for all future days. 
Figure \ref{fig:impact} (bottom) shows how a week after a \IT{} session, almost 30\% more students in the control group drop out compared to those who had \IT{} sessions. 
We believe this significant effect can be attributed to the proven benefits of 1:1 help and also the motivational boost that comes from the personalised attention from a course staff. These are both aspects that are often lacking in traditional MOOCs.

We caveat this result with one confound: people's willingness to accept an offer for 1:1 help might be correlated with their excitement for the course, which in turn affects likelihood of dropout. The fact that the differences in Figure \ref{fig:impact} are not significant for days before the session mitigates this, but it could have a small impact.

\subsection{\IT{}
captures  volunteer teacher time that would otherwise go to waste}

\begin{figure}[h]
    \centering
    \hspace{1mm}
    \includegraphics[width=0.99\linewidth]{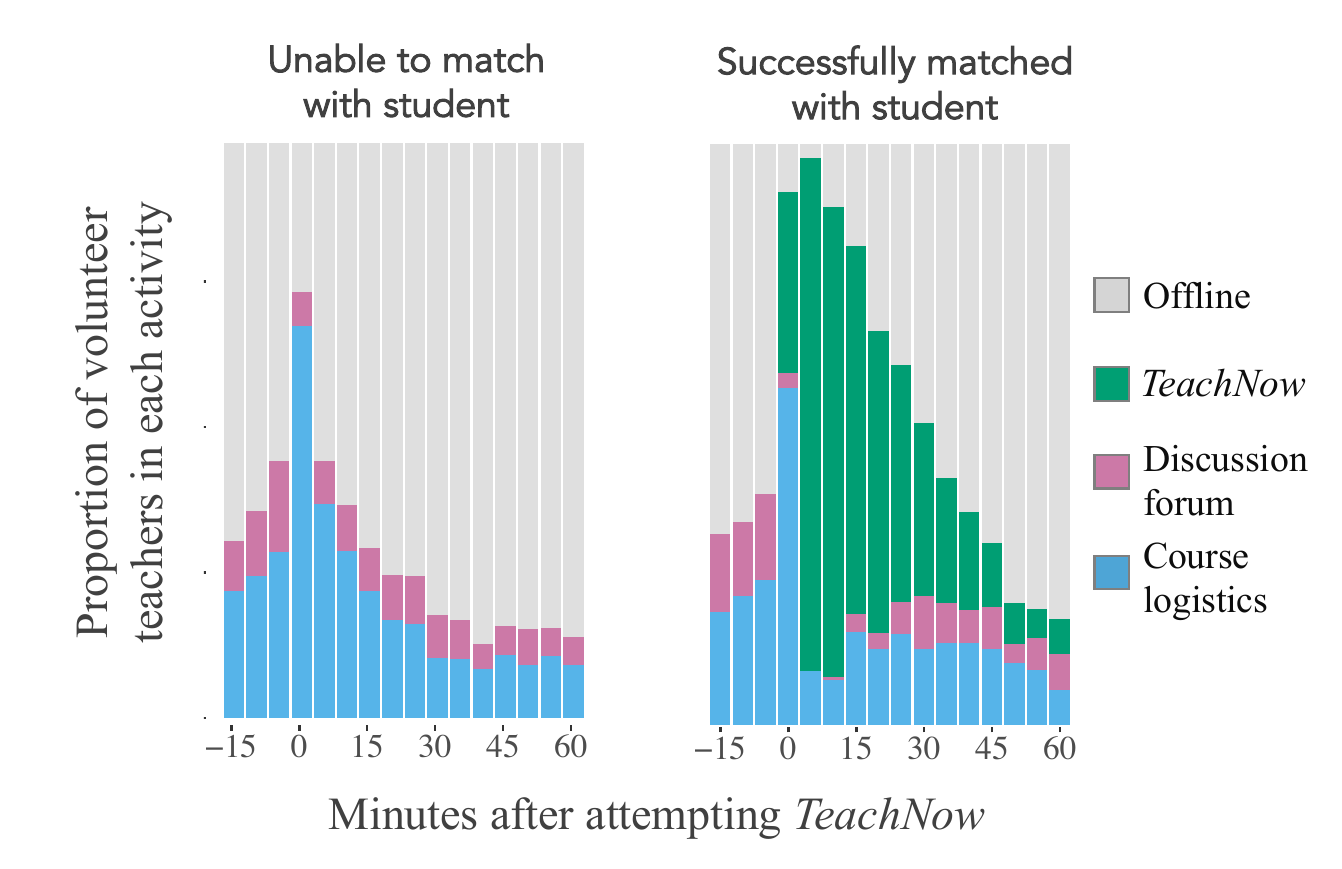}
    \caption{\IT{} captures volunteer time that otherwise goes to waste. Most volunteers who were not able to find a student to help immediately went offline instead of doing other teaching activities.}
    \label{fig:capture_value}
\end{figure}

One of the key ideas behind \IT{} was to allow teachers to use fragmented blocks of free time to help students. Without our system, we hypothesised that volunteers wouldn't have a meaningful alternative activity to pursue during this time.


In order to test this, we looked at instances where a volunteer attempted to do a \IT{} session, but wasn't able to due to random chance. This would occur when several students happened to not want help in that moment.
In such cases, we know the volunteer had 20 minutes of free time to spare, since they attempted a \IT{} session. How would they spend those 20 minutes instead? Would they go on to do some other useful activity like answering questions on the discussion forum or would they immediately go offline? The latter would signal that they were only willing to spend their time on 1:1 help for students. 


Figure \ref{fig:capture_value} explores exactly this question. The $x$-axis represents the minutes leading up to and after a teacher attempted to use \IT{}. The group on the left was unable to match with a student due to random chance, while the group on the right found a successful match. Each bar represents the relative proportion of what volunteer teachers were doing at that moment in time, as measured by their webpage location in our logs.

Both groups look identical in behaviour until the moment they attempt to use \IT{}. For the group able to find a student, the next 15 to 45 minutes are spent providing 1:1 help. For the other group however, we see a majority of teachers going offline instead of doing an alternative activity.
The difference between the green bars on the right and the gray bars on the left is a measure of valuable volunteer time being captured by \IT{}. This finding suggests that \IT{} adds notable value by making use of volunteer teacher time that would otherwise go unseen by students. 

\subsection{Teachers and students find \IT{} valuable, gratifying, and meaningful}
 
\subsubsection{\textbf{Teacher Perspectives}}
We conducted surveys and several interviews with teachers who used \IT{} to understand how they viewed this feature. We found that 73\% of volunteer teachers who used \IT{} rated it among their most enjoyable and impactful activities in the course, and 87\% of them said they would recommend \IT{} to their peers.

We also found that \IT{} fostered teacher development through opportunities for recurrent, `on-demand' teaching experiences. This unique feature presented a novel way for volunteers to improve their teaching skills with real students. 72\% of volunteers ranked \IT{} as the most or second most impactful course activity for their professional development.
We share some of the perspectives on \IT{} that we heard from volunteer teachers in our interviews:
\begin{quote}
``\textit{Watching instant joy and happy reaction students gave after having live help was very satisfying.}''
\end{quote}
\vspace{1mm}
\begin{quote}
``\textit{It felt like I was guaranteed to be helping someone, not just hoping that my fairly quiet section members were getting something out of the section meeting.}
''
\end{quote}
\vspace{1mm}
\begin{quote}
``\textit{I felt that I was impacting the student directly, and seeing how they were learning to solve issues with my support.}
''
\end{quote}


\subsubsection{\textbf{Student Perspectives}} Nearly 80\% of students who had an \IT{} session sent a thank you to their volunteer teacher afterwards, and 68\% of those students felt compelled to also write a personalised message to the teacher. 
The messages talk about how the \IT{} session came just at the right time and helped students feel motivated, supported, and connected to the community:

\begin{quote}
``\textit{I was verge on loosing hope of learning Python. This session gave me the necessary encouragement. You don't know how impactful that was!}" 
\end{quote}
\vspace{1mm}
\begin{quote}
``\textit{It has kept me pushing, moving forward, and it all starting to come together for me. Thank you so much for everything. }". 
\end{quote}
\vspace{1mm}
\begin{quote}``\textit{It's quite encouraging to see that I won't be left behind. I appreciate your check in.}``
\end{quote}

Together, these findings suggest that \IT{} is a mutually beneficial tool, providing a unique platform for ongoing, on-demand teaching moments. This environment enables teachers to evolve in their practice while students benefit from the 1:1 support.

\section{Limitations}\label{sec:lims}


\quad \textbf{\textit{Intelligent matching strategies.}}
Occasionally, our system matched volunteers with students who did not actually need help but just wanted to talk. This was frustrating for volunteers, as many were eager to aid students who were struggling. This expectation mismatch could be addressed with a more nuanced approach to matching. For example, we tried a strategy where volunteers could explicitly ask to match with students who were most behind in the course. A related issue was the student acceptance rate of matches, which hovered around 24\%. Intelligently identifying students to offer help to which maximise match success is an exciting future direction.

\textbf{\textit{Expanding \IT{} to other MOOCs.}}
While we believe \IT{} is a general approach that can be used in any MOOC, there are several challenges. Our deployment was made easier by already having a vetted set of volunteer teachers. Generalising to other MOOCs would require a way to onboard these volunteers, ensure safety, and coordinate expectations.
Additionally, there is a question of the scale necessary for \IT{} to be viable. For example, with fewer students, teachers might not be able to spontaneously find someone to help. Instead we might require our system to identify periods of high demand and encourage teachers to check during those times. 

\textbf{\textit{Technological and social limitations.}}
Our \IT{} sessions required students to have a stable internet connection for video calls. Although we provided  a chat-only mode for those with bandwidth constraints, our observations indicate that these text-only sessions were not as successful. In addition to technological limitations, there are social aspects to consider. \IT{} rests on a student's willingness to meet one-on-one with a volunteer teacher. This is effectively an online interaction with an unfamiliar person, which may be intimidating for students from certain demographics.

\section{Discussion}

Loneliness and isolation are one of the most pressing challenges with online education \cite{Mizani2022Loneliness}. MOOCs in particular provide little opportunity to form social bonds and a sense of community inherent to traditional educational environments.  
We believe \IT{} is a promising approach to bridging this gap. The one-on-one nature of the system facilitates cross-continent connection with strangers who have a shared love of learning and teaching. This is a powerful social activity, and a powerful driver behind its effectiveness on student retention.

Our design of \IT{} connects volunteer teachers to students in MOOCs. An exciting future direction is to also allow \textbf{\textit{students}} to take on teaching roles through \IT{}. Near-peers have been shown to be highly effective tutors \cite{whitman1988peer}. Identifying promising students to take on teaching roles through \IT{} would increase the number of people who can be helped, while also reinforcing the understanding of the student who is teaching \cite{malik2024AVT}. 


Another exciting feature of \IT{} is as a training ground for infinite, repeatable teaching practice. While the primary focus of MOOCs is on student learning, with \IT{}, teachers can also benefit from the real teaching practice they get by providing 1:1 help to students from across the globe. 

As a final observation, the introduction of \IT{} into a global MOOC is a powerful way to facilitate cross-cultural exchange and learning. The \IT{} system is more than a one-on-one help tool; it enables students and teachers to engage with peers from various cultural backgrounds,  promoting much-needed global awareness and cultural learning.

\section{Conclusion}
We present \IT{} as an exciting path forward for scaling 1:1 help to massive online courses using volunteer teachers. We experimentally show the enormous benefits of \IT{} in increasing student retention, making effective use of volunteer teacher time, and fostering a sense of belonging and community in the traditionally isolated virtual world. We believe human-centered real-time learner support in MOOC settings, with further refinement, has the potential to revolutionise the way learning is conducted in massive online courses.

\begin{acks}
We want to thank TJ Jefferson for his help on establishing a collaborative IDE and Jason Ford for setting up collaborative video. We also thank Sierra Wang for her meticuluous data collection, some of which we used for our results. 
\end{acks}

\clearpage
\bibliographystyle{ACM-Reference-Format}
\balance
\bibliography{ref}


\end{document}